\documentclass[pra,twocolumn,floatfix,preprintnumbers,showpacs,superscriptaddress]{revtex4-1}
\usepackage[english]{babel}
\usepackage[latin1]{inputenc}
\usepackage{enumerate}
\usepackage{amsfonts}
\usepackage{amssymb}
\usepackage{amsmath,amsbsy}
\usepackage{dcolumn}
\usepackage{algpseudocode}
\usepackage{bm}
\usepackage{graphicx, graphics}
\usepackage{color}
\usepackage[breaklinks=true, pdftex]{hyperref}
\usepackage{multirow}

\usepackage{multibib}

\hypersetup{
    colorlinks=true,
    linkcolor=[rgb]{0.0,0.2,0.8},
    citecolor=[rgb]{1.0,0.0,0.0},
    filecolor=magenta,
    urlcolor=[rgb]{0.0,0.3,0.7}
}
\newcommand{\pap}[1]{\left( #1 \right)}
\newcommand{\pas}[1]{\left[#1 \right]}

\newcommand{\ket}[1]{\left\vert #1 \right\rangle}

\newcommand{\ee}{\mathrm{e}}
\newcommand{\ii}{\dot{\iota}}

\newcommand{\eps}{\epsilon}

\begin{document}
	\title{Universal two-time correlations, out-of-time-ordered correlators and Leggett-Garg inequality violation by edge Majorana fermion qubits}
	\author{F. J. Gómez-Ruiz}
		\email{ fj.gomez34@uniandes.edu.co}
		\affiliation{Departamento de F{\'i}sica, Universidad de los Andes, A.A. 4976, Bogot\'a D. C., Colombia.}
		\affiliation{Department of Physics, University of Massachusetts, Boston, MA 02125, USA}
	\author{J. J. Mendoza-Arenas}
		\affiliation{Departamento de F{\'i}sica, Universidad de los Andes, A.A. 4976, Bogot\'a D. C., Colombia.}
	\author{F. J. Rodr{\'i}guez}
		\affiliation{Departamento de F{\'i}sica, Universidad de los Andes, A.A. 4976, Bogot\'a D. C., Colombia.}
	\author{C. Tejedor}
		\affiliation{Departamento de F{\'i}sica Te{\'o}rica de la Materia Condensada and Condensed Matter Physics Center (IFIMAC), Universidad Aut{\'o}noma de Madrid, 28049, Spain.}
        \author{L. Quiroga}
        		\affiliation{Departamento de F{\'i}sica, Universidad de los Andes, A.A. 4976, Bogot\'a D. C., Colombia.}
	\date{\today}
\begin{abstract}
 In the present work we propose that two-time correlations of Majorana edge localized fermions constitute a novel and versatile toolbox for assessing the topological phases of 1D open lattices. Using analytical and numerical calculations on the Kitaev model, we uncover universal relationships between the decay of the short-time correlations and a particular family of out-of-time-ordered correlators, which provide direct experimental alternatives to the quantitative analysis of the system regime, either normal or topological. Furthermore we show that the saturation of two-time correlations possesses features of an order parameter. Finally, we find that violations of Leggett-Garg inequalities can indicate the topological-normal phase transition by looking at different qubits formed by pairing local and non-local edge Majorana fermions.
		
		\end{abstract}
		\maketitle
\section{Introduction} \label{intro}
 In the last few years, the development of new quantum devices has fuelled the search for novel materials and control mechanisms to engineer unprecedented technologies. Along this path, topological systems have been identified as robust entities with potential applications in quantum computation and information processing due to their unusual braiding properties~\cite{alisea1,dassarma1,Sau}. Candidates for topological qubits include chains of magnetic atoms on top of a superconducting surface~\cite{Nadj}, hybrid systems between $s-$wave superconductors and topological insulators~\cite{HaltermanPRB}, $p-$wave superconductors~\cite{TanakaPRB}, fractional quantum Hall systems~\cite{MooreNP} and 1D semiconductor-superconductor heterostructure based quantum wires~\cite{RomanPRL,Mourik1003,Das}. Notably, the latter have aroused great interest given their high experimental accessibility and controllability~\cite{DumitrescuPRB}. In addition, edge-localized Majorana zero modes, expected to be robust against dephasing and dissipation~\cite{elliott,dassarma2,AguadoR,Albrecht}, have been predicted to exist in these systems. The search of new topological configurations allowing for Majorana zero modes has also been extended to Josephson junction based nanostructures~\cite{KuertenPRB,KalcheimPRB,AliPRB,MohaPRB,Alidoust}. \\
\\
Concurrently with the chase of novel materials is the search for experimentally-accessible properties to identify their truly nonclassical features, such as topological quantum phases. A large number of protocols have been proposed to this end, and a particularly important subset are those based on spatial non-local correlations as embodied in Bell inequalities~\cite{Brunner_RMP2014,drummond2014prb,dassarma2}.
More recently there has been a surge of theoretical and experimental interest in using temporal correlations instead for similar purposes, since in some scenarios nonlocal measurements are quite challenging. Thus local measurements such as two-time correlations (TTC) can be used to gain further access to the underlying physics~\cite{Gessner2014epl,Gessner2014Nat}.\\
\\
Here we consider an extension of that interest to assess the interplay between time correlations and nonlocal quantum objects in Majorana fermion chains, a situation different from any other previously considered, by focusing mainly on the Kitaev chain~\cite{kitaev}. In particular we address the open question of detecting true quantum temporal correlations in a topological quantum phase. Correlations for two types of objects are to be explored: {\bf (i)} Local Dirac fermions formed by pairing two Majorana fermions on the same edge site, and {\bf (ii)} Non-local Dirac fermions coming from the pairing of Majorana fermions located at the two opposed edge sites of the chain.  In this way we will address the pivotal role that TTCs play for detecting large memory effects of local and non-local Majorana edge qubits.\\
\\
 Specifically we will show how the longtime limit of several boundary TTCs possesses features of an order parameter, providing information on the quantum phase transitions of the Majorana fermion system. Moreover, for the purposes of the present work, TTC can be used to assess the quantumness of a system, in a form similar to spatial correlations do through Bell inequalities. Namely, combinations of TTC allow for testing Leggett-Garg inequalities (LGIs)~\cite{leggett,Leggett2,emary}. These inequalities are satisfied in macroscopic classical systems, characterized by macrorealism (a system's property is well defined at every time regardless of being observed or not) and noninvasive measurability (a system is unaffected by measurements). Their violation indicates the existence of macroscopic quantum coherence.\\
\\
Not only there has been an intense search for experimental schemes in which LGI violations can be observed~\cite{palacios,goggin,knee,athalye,waldherr,dressel,bell_leggett_garg_2016,huffman2017pra}, but also several applications for them have been proposed, including identification of order-disorder quantum phase transitions in many-body systems~\cite{gomez-ruiz} and characterization of quantum transport~\cite{lambert2010prl}. Indeed it is also interesting to extend the range of LGI violation features as a detection tool of topological phase transitions. Along this line, our results provide a first step for understanding the link between correlations in space and time domains in a concrete topological condensed-matter set up. Moreover, we stress that all of our results remain still valid for an edge spin in the transverse field Ising open chain, by applying a Jordan-Wigner transformation to the open Kitaev chain model.\\
\\
 The paper is organized as follows. Section~\ref{sec_2} gathers up a brief review of specific Majorana fermion chains (in the language of the 1D Kitaev model), its exact diagonalization and theoretical aspects of Majorana qubit two-time correlations. In Section~\ref{sec_5} numerical results of Majorana qubit-TTCs behavior for short-, intermediate- and long-time regimes are discussed. In Section~\ref{sec_6} a brief recap of LGI is brought out, with focus on the intermediate time regime.  Both analytical and numerical results are provided and contrasted whenever possible. Section~\ref{sec_7} is devoted to discuss possible experimental implementations where TTC and LGI behaviors of Majorana-based qubits could be tested. Finally, in Section~\ref{sec_8} we present a summary of this work.

\section{Theoretical framework}\label{sec_2}
\subsection{Majorana fermion chain}
We focus on a concrete realization of a Majorana fermion chain in terms of the Kitaev model~\cite{kitaev}. It is described by the Hamiltonian:
\begingroup\makeatletter\def\f@size{7.85}\check@mathfonts
\def\maketag@@@#1{\hbox{\m@th\large\normalfont#1}}
\begin{equation}\label{Hkitaev}
\begin{split}
\hat{H}=&-\frac{\mu}{2}\sum_{j=1}^{N} \left(2\hat{n}_{j}-1\right)\\
&-\omega\sum_{j=1}^{N-1}\left(\hat{c}_{j}^{\dagger}\hat{c}_{j+1}+\hat{c}_{j+1}^{\dagger}\hat{c}_{j}\right)+\Delta \sum_{j=1}^{N-1}\left(\hat{c}_{j}\hat{c}_{j+1}+\hat{c}_{j+1}^{\dagger}\hat{c}_{j}^{\dagger}\right),
\end{split}
\end{equation}
\endgroup
representing a system of non-interacting  spinless fermions on an open end chain of $N$ sites labeled by $j=1,\ldots, N$. The single site fermion occupation operator is denoted by $\hat{n}_{j}=\hat{c}_{j}^{\dagger}\hat{c}_{j}$, the chemical potential is $\mu$, taken as uniform along the chain, $\omega$ is the hopping amplitude between nearest-neighbor sites (we assume $\omega\geq0$ without loss of generality because the case with $\omega\leq0$ can be obtained by a unitary transformation: $\hat{c}_{j}\to-\ii\left(-1\right)^{j}\hat{c}_{j}$) and $\Delta$ is the $p-$wave paring gap, which is assumed to be real and $\Delta\geq 0$ (the case $\Delta\leq0$ can be obtained by transformation $\hat{c}_{j}\to \ii\,\hat{c}_{j}$ for all $j$). This model captures the physics of a 1-D topological superconductor with a phase transition between topological and nontopological (trivial) phases at $\mu=2\Delta$, for $\Delta=\omega$. Notice that for this symmetric hopping-pairing Kitaev Hamiltonian, i.e. $\omega=\Delta$, a Jordan-Wigner transformation leads directly into the transverse field Ising model~\cite{leenjp16}. Thus, from now on we will refer as Majorana fermion chain either the Kitaev chain or the transverse field Ising model.\\

Let us introduce Majorana operators $\hat{\gamma}_{j}$ to express the real space spinless fermion annihilation and creation operators, as:
\begin{equation}\label{Majo_op}
\hat{c}_j=\frac{1}{2}\left ( \hat{\gamma}_{2j-1}+\ii\hat{\gamma}_{2j} \right ), \qquad \hat{c}_j^{\dagger}=\frac{1}{2}\left ( \hat{\gamma}_{2j-1}-\ii\hat{\gamma}_{2j} \right ).
\end{equation}
These are Hermitian operators $(\hat{\gamma}_j=\hat{\gamma}_j^{\dagger})$, satisfy the property $\left(\hat{\gamma}_{j}\right)^{2}=\left(\hat{\gamma}_{i}^{\dagger}\right)^{2}=1$, and obey the modified anticommutation relations $\lbrace\hat{\gamma}_{i},\hat{\gamma}_{j}\rbrace=2\delta_{i,j}$, with $i,j=1,\dots,2N$. From the definition of Majorana operators~\eqref{Majo_op} it is evident that for each spinless fermion on site $j$, two Majorana fermions are assigned to that site, which are denoted by $\hat{\gamma}_{2j-1}$ and $\hat{\gamma}_{2j}$.  They allow the Kitaev Hamiltonian in Eq.~\eqref{Hkitaev} to be written in the equivalent form:
\begingroup\makeatletter\def\f@size{8.5}\check@mathfonts
\def\maketag@@@#1{\hbox{\m@th\large\normalfont#1}}
\begin{equation}\label{Hmajo1}
\begin{split}
\hat{H}=&-\ii \frac{\mu}{2}\sum_{j=1}^{N}\hat{\gamma}_{2j-1}\hat{\gamma}_{2j}\\
&+\frac{\ii}{2}\sum_{j=1}^{N-1}\left[(\omega+\Delta)\;\hat{\gamma}_{2j}\hat{\gamma}_{2j+1}-(\omega-\Delta)\;\hat{\gamma}_{2j-1}\hat{\gamma}_{2j+2}\right].
\end{split}
\end{equation}
\endgroup

In order to put the Majorana fermion Hamiltonian in Eq.~\eqref{Hkitaev} (or equivalently in Eq.~\eqref{Hmajo1}) in diagonal form, a standard Bogoliubov transformation (see Supplemental Material (SM)~\cite{*[{See the Supplemental Material at }] [{for details of the calculations and derivations.}] SM_Gomez}) is performed:

\begin{equation}\label{BGT}
\begin{split}
\hat{c}_{j}^{\dagger} &= \sum_{k=1}^{N} \left ( u_{2k,j} \hat{d}_{k} + v_{2k,j}\hat{d}^{\dagger}_{k}\right ) , \\
\hat{c}_{j} &= \sum_{k=1}^{N} \left ( u_{2k,j} \hat{d}^{\dagger}_{k} + v_{2k,j}\hat{d}_{k}\right ),
\end{split}
\end{equation}
where $k$ denotes a single fermion mode, $u_{2k,j}$ and $v_{2k,j}$ are real numbers, and the canonical fermion anticomutation relations for the new operators $\hat{d}_{k}$, $\hat{d}^{\dagger}_{k}$ remain true, that is $\left\{\hat{d}_{k},\hat{d}^{\dagger}_{k'}\right\}=\delta_{k,k'}$, $\left\{\hat{d}^{\dagger}_{k},\hat{d}^{\dagger}_{k'}\right\} =\left\{\hat{d}_{k},\hat{d}_{k'}\right\}=0$.
Thus the exact diagonalization of the Kitaev Hamiltonian in Eq.~\eqref{Hkitaev}, in terms of the new independent fermion mode operators $\hat{d} \pap{\hat{d}^{\dagger}}$, leads to:
\begin{equation}\label{H_num}
\hat{H}=\sum_{k=1}^{N}\eps_{k}\pas{\hat{d}^{\dagger}_{k}\hat{d}_{k}-\frac{1}{2}},
\end{equation}
where the new fermion mode energies $\eps_{k}\geq 0$ are to be numerically calculated for a Kitaev chain with open ends (although analytical exact results may be found in some cases, see~\cite{narozhny}). The matrix representation of Eq.~\eqref{H_num} is explicitly written in the SM~\cite{SM_Gomez}.
\subsection{Two-time correlations}
The key quantity of interest in the present work is the symmetrized TTC, $\mathcal{C}\pap{t_{1},t_{2}}$, as given by the expression~\cite{gomez-ruiz}
\begin{equation}\label{eq:Correl}
\mathcal{C}\pap{t_{1},t_{2}}=\frac{1}{2}\big\langle \lbrace \hat{Q}\pap{t_2},\hat{Q}\pap{t_{1}}\rbrace \big\rangle,
\end{equation}	
where $\hat{Q}$ denotes a single qubit operator (a dichotomic observable, i.e. with eigenvalues $q=\pm 1$) to be specified later, $\lbrace \hat{Q},\hat{Q}^{\prime}\rbrace$ is an anticommutator and $\hat{Q}(t_{n})$ the qubit operator at time $t_{n}$. Since the TTC is to be evaluated for stationary states, it does not depend on the individual times $t_1$ and $t_2$ but only on their difference $t=t_2-t_1$, leading simply to $\mathcal{C}\pap{t_1,t_2}=\mathcal{C}\pap{t,0}=\mathcal{C}\pap{t}$.\\
\\
In the following subsections, we present an analytical approach for the evaluation of TTC for single and double Majorana qubits, together with extensive supporting numerical data. In particular in Sec.~\ref{sec_3} we obtain the early-time TTC behavior for different local and non-local qubits, demonstrating that for the local single and double Majorana qubit case universal features (independent of the chain quantum state itself as well as chain size) can be identified providing local information about the global systems many-body quantum phase. We also link the TTCs short-time evolution to a special kind of recently established many-time correlators, the so called out-of-time-ordered correlators, which are gaining growing interest. Moreover, in Sec.~\ref{sec_4} we provide analytical results for single- and double-Majorana TTCs for arbitrary times, showing how the former serves as a test of topological criticality by directly indicating the existence of edge localized zero energy modes. These results are then fully evaluated numerically in Sec.~\ref{sec_5}.

\subsection{TTC short-time behavior and out-of-time-ordered correlation function} \label{sec_3}

In order to assess the sensitivity of TTC for detecting quantum phase transitions by looking at a single local site, we connect the short time TTC behavior to a second-order expansion with out-of-time-ordered correlation function $\mathcal{T}\pap{t}=\langle \hat{O}^{\dagger}_{1}\pap{t}\hat{O}^{\dagger}_{2}\pap{0}\hat{O}_{1}\pap{t}\hat{O}_{2}\pap{0} \rangle$~\cite{swingle2016pra,shen2017prb,garttner2017nat,li2017prx}. Let us expand up to second order in time the TTC as given by Eq.~\eqref{eq:Correl}, yielding to
\begin{equation}\label{otoc1}
\begin{split}
\mathcal{C}\pap{t}&=\frac{1}{2}\langle \lbrace \ee^{\ii \hat{H}t}\hat{Q}\ee^{-\ii \hat{H}t},\hat{Q}\rbrace \rangle\\
&\simeq 1-\frac{t^2}{2}\langle -[ \hat{H},\hat{Q} ]^2\rangle+{\cal O}(t^4)
\end{split}
\end{equation}
Note that the second line in the last equation holds for any single-site qubit observable $\hat{Q}$ such that $\hat{Q}^2=\hat{1}$, evolving under the action of an arbitrary (local or global) Hamiltonian $\hat{H}$. Moreover, and most interestingly, the first line in Eq.~\eqref{otoc1} is nothing but the real part of the $\mathcal{T}(t)$ corresponding to a hermitian single qubit operator $\hat{O}_2=\hat{Q}$ and the unitary operator $\hat{O}_{1}(t)=e^{-iHt}$. \\
\\
First, let us consider the TTC for a single edge Majorana fermion $j=1$, i.e. $\hat{Q}=\hat{\gamma}_{1}$. By resorting to Eq.~\eqref{Hmajo1} it is easy to check that $[ \hat{H},\hat{\gamma}_{1} ]^2=-\mu^2$, a scalar quantity, thus producing for the real part of the corresponding $\mathcal{T}(t)$ the simple and universal result $\langle -[ \hat{H},\hat{\gamma}_{1} ]^2\rangle=\mu^2$, valid for any Majorana fermion chain pure state $\ket{\psi_{K}}$ or mixed state $\hat{\rho}_K$. Note that via the Jordan-Wigner transformation, this qubit operator corresponds to $\hat{\sigma}_{1}^{x}$ for the transverse field Ising model, i.e. $\hat{\gamma}_{1}=\hat{\sigma}_{1}^{x}$, the $x$-spin operator of an edge chain site. Consequently we rewrite the TTC in Eq.~\eqref{otoc1} as $\mathcal{C}_{1}^{(x)}\pap{t}$,
\begin{equation}\label{otoc2}
\begin{split}
\mathcal{C}_{1}^{(x)}\pap{t}&=\frac{1}{2}\big\langle \lbrace \hat{\gamma}_1\pap{t},\hat{\gamma}_1\rbrace \big\rangle\\
&\simeq 1-\frac{\mu^2}{2}t^2+{\cal O}(t^4).
\end{split}
\end{equation}

As a second case, we consider a two-Majorana edge qubit such as $\hat{Q}=2\hat{n}_1-\hat{1}=-\ii\,\hat{\gamma}_{1}\hat{\gamma}_{2}$. This qubit corresponds, via the Jordan-Wigner transformation, to the $\hat{\sigma}_{1}^{z}$ edge spin operator for the transverse field Ising model, i.e. $-\ii\,\hat{\gamma}_{1}\hat{\gamma}_{2}=\hat{\sigma}_{1}^{z}$. Now it is straightforward to show that $[ \hat{H},-\ii\,\hat{\gamma}_{1}\hat{\gamma}_{2} ]^2=-4\Delta^2$, again a scalar quantity and hence producing a result valid for any Majorana fermion chain pure state $\ket{\psi_K}$ or mixed state $\hat{\rho}_K$. Thus the second derivative of the real part of the $\mathcal{T}(t)$ reduces to the universal value $\langle -[ \hat{H},-\ii\,\hat{\gamma}_{1}\hat{\gamma}_{2} ]^2\rangle=4\Delta^2$ and consequently the short time expression for $\mathcal{C}_{1}^{(z)}\pap{t}$ becomes
\begin{equation}\label{otoc3}
\begin{split}
\mathcal{C}_{1}^{(z)}\pap{t}&=-\frac{1}{2}\big\langle \lbrace \hat{\gamma}_1\pap{t}\hat{\gamma}_2\pap{t},\hat{\gamma}_1\hat{\gamma}_2\rbrace \big\rangle \\
&\simeq 1-2\Delta^2t^2+{\cal O}(t^4).
\end{split}
\end{equation}
\begin{figure}[t!]
  \includegraphics[width=0.45 \textwidth]{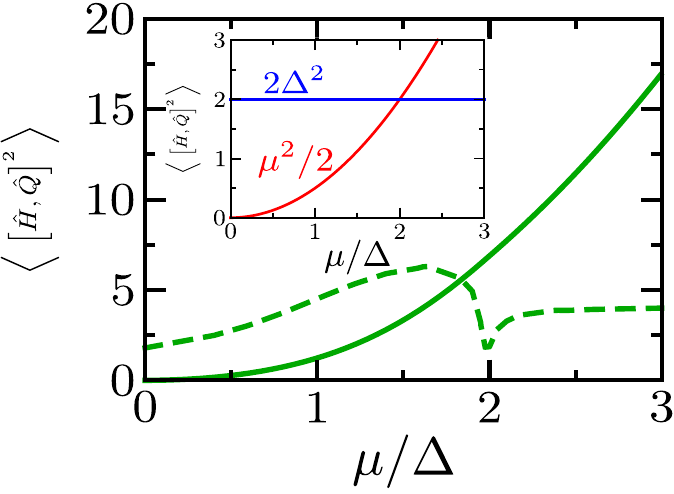}
  \caption{(color online)  Short-time curvatures of the different edge TTCs as a function of $\mu/\Delta$. Main panel: Curvature for the non-local two-Majorana qubit $\mathcal{C}_{1,N}\pap{t}$ (green, solid line),  and its second derivative with respect to $\mu$ (green, dashed line). The latter presents a clear dip at the topological quantum critical point. Inset: TTC initial curvature for local Majorana qubits, $\mathcal{C}_{1}^{x}\pap{t}$ (red line) and $\mathcal{C}_{1}^{z}\pap{t}$ (blue line), showing a crossing just at the critical point.}
  \label{fig_1}
\end{figure}

 As a third case, we analyze the short-time behavior of the non-local Dirac fermion formed by coupling two Majorana operators located at the two edges of the chain, $\hat{Q}_{1,N}=\ii\,\hat{\gamma}_{1}\hat{\gamma}_{2N}$. The expansion of the corresponding TTC leads to
\begingroup\makeatletter\def\f@size{9.5}\check@mathfonts
\def\maketag@@@#1{\hbox{\m@th\large\normalfont#1}}
\begin{equation}\label{otoc20}
\begin{split}
\mathcal{C}_{1,N}\pap{t}&= -\frac{1}{2}\big\langle \lbrace \hat{\gamma}_1\pap{t}\hat{\gamma}_{2N}\pap{t},\hat{\gamma}_1\hat{\gamma}_{2N}\rbrace \big\rangle\\
&\simeq 1-\mu^2\left ( 1- \langle \hat{\gamma}_1\hat{\gamma}_2\hat{\gamma}_{2N-1}\hat{\gamma}_{2N}\rangle \right )\,t^2+{\cal O}(t^4).
\end{split}
\end{equation}
\endgroup
It is evident that this TTC features a non-universal short-time evolution, given that it depends on the specific quantum state of the Majorana fermion chain. This is indicated by the expected value of the four Majorana operator term $\langle \hat{\gamma}_1\hat{\gamma}_2\hat{\gamma}_{2N-1}\hat{\gamma}_{2N}\rangle=\langle \hat{\gamma}_1\hat{\gamma}_2\rangle\langle \hat{\gamma}_{2N-1}\hat{\gamma}_{2N}\rangle+\langle \hat{\gamma}_1\hat{\gamma}_{2N}\rangle\langle \hat{\gamma}_2\hat{\gamma}_{2N-1}\rangle$, which for a sufficiently long it can be approximated to $\langle \hat{\gamma}_1\hat{\gamma}_2\hat{\gamma}_{2N-1}\hat{\gamma}_{2N}\rangle\simeq \langle \hat{\gamma}_1\hat{\gamma}_2\rangle^2$ in the ground state.\\
\\
In Fig.~\ref{fig_1} the short-time curvature (second time derivative) of the edge TTCs corresponding to single- and double-Majorana fermions is depicted as a function of $\mu/\Delta$. In the main panel the non-local case of $\mathcal{C}_{1,N}\pap{t}$ is plotted (solid line), while in the inset those of $\mathcal{C}_{1}^{(x)}\pap{t}$ and $\mathcal{C}_{1}^{(z)}\pap{t}$ are depicted. Clearly, by comparing Eqs.~\eqref{otoc2}-\eqref{otoc3}, an universal crossing of initial TTC curvatures occurs for $\mu=2\Delta$, which signals the critical point for the topological-trivial phase transition in the Kitaev model, or equivalently for the ferromagnetic-paramagnetic transition in the transverse field Ising model. This remarkable universal behavior, i.e. the independence from the Majorana fermion chain quantum state, holds true only for the edge sites of both these models as realized by the Kitaev chain and transverse field Ising systems.  By contrast, the non-local $\mathcal{C}_{1,N}\pap{t}$ shows a non-universal behavior depending on the specific quantum state of the Majorana fermion chain. The results plotted in the main panel of Fig.~\ref{fig_1} have been obtained numerically, as explained below, for a Kitaev chain in the ground state. In the same panel the second derivative of the curvature with respect to $\mu$ is also plotted (dashed line), which clearly presents a dip at the critical point $\mu/\Delta=2$. Thus, we observe that the early-time correlators, both with universal and non-universal behavior, are sensitive to the topological phase transition.


\subsection{General two-time correlation behavior of Majorana qubits} \label{sec_4}

Having established the relevance of TTCs and a related family of out-of-time-ordered correlations for edge sites in the Majorana fermion chain, we proceed to explore the TTC behavior for qubits formed by any combination between edge and/or bulk sites, for arbitrary times. By developing the Majorana qubits in terms of Bogoliubov operators (see SM~\cite{SM_Gomez}) we proceed to express both the single- and double-Majorana TTC in convenient forms for numerical analysis. We follow the same notation for Bogoliubov coefficients used in Ref.~\cite{Dmytruk}. As we discuss below this numerical procedure is essential to further progress, except in special cases  for $\mathcal{C}_{1}^{(x)}\pap{t}$ where an exact closed form has been obtained.
\subsubsection{Single-Majorana edge two-time correlations}
First we note that the $\mathcal{C}_{1}^{(x)}(t)$-TTC admits an universal exact closed expression for arbitrary pure or mixed quantum states of the Majorana fermion chain (see SM~\cite{SM_Gomez}).  We found that
\begin{equation}\label{naraya1}
\mathcal{C}_{1}^{(x)}\pap{t}=\sum_{m=0}^{\infty}\frac{(-1)^m}{(2m)!}\left ( 2\Delta t \right )^{2m}\mathcal{N}_m(u^2),
\end{equation}
where $u=\frac{\mu}{2\Delta}$, and $\mathcal{N}_m(x)=\sum_{n=1}^{m}N_{m,n}x^{n}$ are the well-known Narayana polynomials, which involve the Narayana numbers $N_{m,n}=\frac{1}{m}\binom {m}{n-1}\binom {m}{n}$~\cite{Kostov,Sun}. Note that the critical point corresponds to $u=1$, for which $\mathcal{N}_m(1)=C_m=\frac{1}{m+1}\binom {2m}{m}$, the most famous Catalan numbers. Importantly Eq.~\eqref{naraya1} can be calculated in a closed form at the critical point $u=1$, yielding to the simple expression
\begin{equation}\label{naraya2}
\mathcal{C}_{1}^{(x)}\pap{t}=\frac{\mathcal{J}_1\left ( 4\Delta t \right )}{2\Delta t}
\end{equation}
in terms of the Bessel function of the first kind $\mathcal{J}_1(z)$. To the best of our knowledge this compact result has passed unnoted in the literature on both Ising and Kitaev models. We emphasize that the expressions given by Eqs.~\eqref{naraya1}-\eqref{naraya2} are always valid and thus they are of universal reach, independently of the pure or mixed state of the Majorana fermion chain. Consequently, they hold true even at infinite temperature.\\
\\
For other values of $u$ such a simple form has yet to be found. However the analytics can be developed further, leading to deeper insights on the general behavior of the TTC. First, Eq.~\eqref{naraya1} allows for establishing a link of $\mathcal{C}_{1}^{(x)}(t)$-TTC on both phases around the critical point $u=1$, which will come in handy afterwards. Since the Narayana polynomials are symmetric, the property $\mathcal{N}_m(\frac{1}{x})=\frac{1}{x^{m+1}}\mathcal{N}_m(x)$ holds. Consequently,
\begin{equation}\label{naraya3}
\mathcal{C}_{1}^{(x)}\pap{t,\frac{1}{u}}=1-\frac{1}{u^2}+\frac{1}{u^2}\,\,\mathcal{C}_{1}^{(x)}\pap{\frac{t}{u},u},
\end{equation}
indicating that the $x$-TTC behaves in one phase (reduced chemical potential $\frac{1}{u}$) as it would do in the complementary phase (reduced chemical potential $u$) but with a scaled time $\frac{t}{u}$.\\
\\
Furthermore, for numerical calculations the time evolution of a single Majorana edge fermion operator, $\hat{\gamma}_{i}(t)=\ee^{\ii \hat{H}t}\hat{\gamma}_{i}(0)\ee^{-\ii \hat{H}t}$, is found to be
\begin{figure*}[t!]
  \includegraphics[width=1 \textwidth]{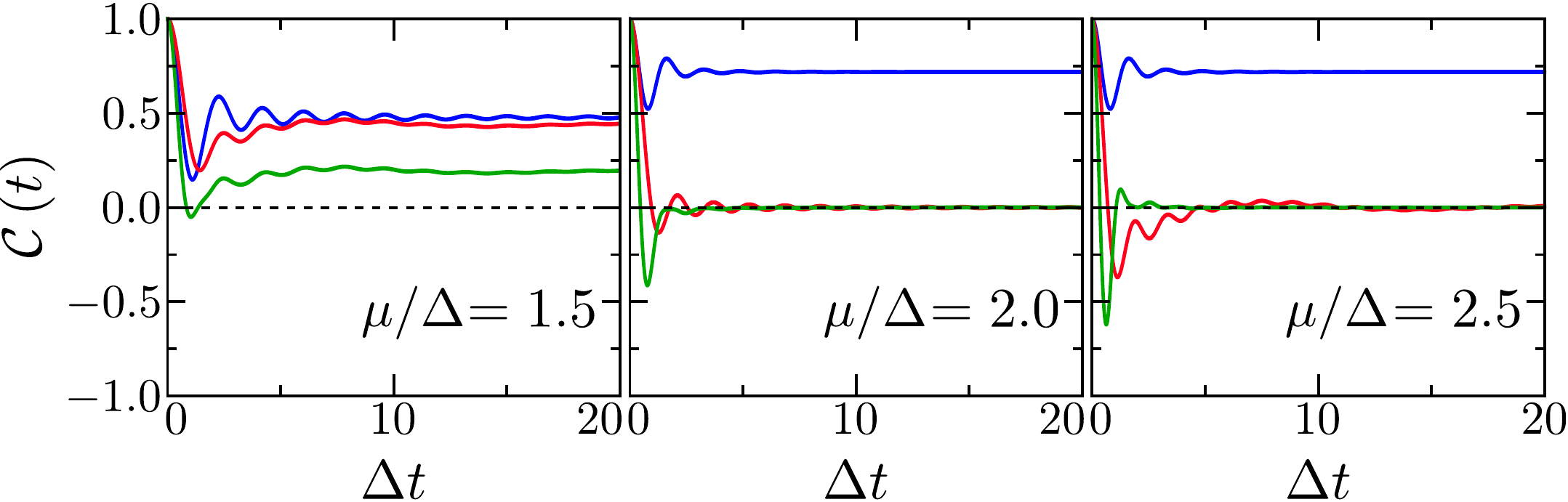}
  \caption{(color online) Edge single- and double-Majorana qubit TTC in the topological phase ($\mu/\Delta=1.5$, left panel), at the transition point ($\mu/\Delta=2$, central panel) and in the non-topological phase ($\mu/\Delta=2.5$, right panel). In all panels the red line depicts the $\mathcal{C}_{1}^{(x)}\left(t\right)$-TTC, the blue line represents the $\mathcal{C}_{1}^{(z)}\left(t\right)$-TTC, while the green line corresponds to $\mathcal{C}_{1,N}\left(t\right)$.}
  \label{fig_2}
\end{figure*}
\begin{equation}\label{otoc4}
\begin{split}
\hat{\gamma}_{2j-1}\pap{t}&=\sum_{m=1}^N\lbrace \hat{\gamma}_{2m-1}\,g_{m,j}^{(+,+)}(t)+\hat{\gamma}_{2m}\,h_{m,j}^{(-,+)}(t) \rbrace\\
\hat{\gamma}_{2j}\pap{t}&=\sum_{m=1}^N\lbrace \hat{\gamma}_{2m}\,g_{m,j}^{(-,-)}(t)-\hat{\gamma}_{2m-1}\,h_{m,j}^{(+,-)}(t) \rbrace,
\end{split}
\end{equation}
where
\begingroup\makeatletter\def\f@size{7.85}\check@mathfonts
\def\maketag@@@#1{\hbox{\m@th\large\normalfont#1}}
\begin{equation}\label{otoc5}
\begin{split}
g_{m,j}^{(\nu,\nu)}(t)&=\sum_{k}{\rm cos}(\epsilon_k\,t)\,(u_{2k,m}+\nu v_{2k,m})\,(u_{2k,j}+\nu v_{2k,j})\\
h_{m,j}^{(\nu,-\nu)}(t)&=\sum_{k}{\rm sin}(\epsilon_k\,t)\,(u_{2k,m}-\nu v_{2k,m})\,(u_{2k,j}+\nu v_{2k,j}),
\end{split}
\end{equation}
\endgroup
with $\nu=+,-$. A direct application of these relations allows us to obtain an analytical expression for the full time evolution of $\mathcal{C}_{1}^{(x)}\pap{t}$, as
\begin{equation}\label{otoc6}
\mathcal{C}_{1}^{(x)}\pap{t}=\sum_{k}{\rm cos}(\epsilon_k\,t)(u_{2k,1}+v_{2k,1})^2,
\end{equation}
where $\langle \gamma_{2i}\gamma_{2j-1}\rangle=-i\sum_{k}\left (u_{2k,i}-v_{2k,i}\right )\left (u_{2k,j}+v_{2k,j}\right )$ and $\langle \gamma_{2i}\gamma_{2j}\rangle=\langle \gamma_{2i-1}\gamma_{2j-1}\rangle=\delta_{i,j}$ have been used. By expanding Eq.~\eqref{otoc6} up to second order in time and comparing it with the universal result quoted in Eq.~\eqref{otoc2} the following identity holds true,
\begin{equation}\label{otoc7}
\sum_{k}\epsilon_k^2\,\left (u_{2k,1}+v_{2k,1}\right )^2=\mu^2,
\end{equation}
which is valid for open Kitaev and transverse field Ising models (with $\mu$ replaced by the transverse magnetic field) of arbitrary chain length. The identity given by Eq.~\eqref{otoc7} provides by itself a consistency check of numerical calculations.\\
\\
Now let us look at the long-time limit of $\mathcal{C}_{1}^{(x)}\pap{t}$ by averaging Eq.~\eqref{otoc6} over a long time period. As the time average of $\cos(\epsilon_k\,t)$ vanishes unless some fermion mode has energy $\epsilon_M=0$, i.e. a zero energy Majorana mode exists  (for which the average is $1$), we can readily assure that for the topological regime
\begin{equation}\label{otoc11}
\lim_{t\to\infty} \mathcal{C}_{1}^{(x)}\pap{t} \simeq (u_{M,1}+v_{M,1})^2=4u_{M,1}^2,
\end{equation}
since $u_{M,1}=v_{M,1}$, i.e. the electron and hole contributions for the zero energy Majorana mode $k=M$ at site $j=1$ are the same. Consequently, we propose that a measurement of the long time saturation value of the edge $\mathcal{C}_{1}^{(x)}$-TTC provides a witness of the topological ($\neq 0$) and non-topological ($=0$) phase transition of the Majorana fermion chain systems, as it probes directly the existence of zero energy modes. Additionally, it gives direct access to the electron-hole weight of such modes.\\
\subsubsection{Two-time correlations of double-Majorana qubits}
We focus now on qubits formed by any pair of Majorana fermions such as $\hat{\gamma}_{2i-1}$ and $\hat{\gamma}_{2j}$; details of the calculations are given in the SM~\cite{SM_Gomez}. We define
\begin{equation}\label{maj1}
\hat{\theta}_{i,j}=\frac{1}{2}\left ( \hat{\gamma}_{2i-1}+\ii\,\hat{\gamma}_{2j} \right ), \,
\hat{\theta}_{i,j}^{\dagger}=\frac{1}{2}\left ( \hat{\gamma}_{2i-1}-\ii\,\hat{\gamma}_{2j} \right).
\end{equation}
Notice that $i=j$ implies that the forming Majorana modes are located on the same physical site, and the Kitaev operators in Eq.~\eqref{Hkitaev} are recovered, i.e $\hat{\theta}_{j,j}=\hat{c}_j$. On the other hand, for $i\neq j$ the Majorana fermions are located on different physical sites. It is easy to check that usual Dirac fermion relations hold true for operators $\hat{\theta}_{i,j}$ and $\hat{\theta}_{i,j}^{\dagger}$ as $\left \{\hat{\theta}_{i,j},\hat{\theta}_{i,j}^{\dagger}\right \}=1$, $\left\{\hat{\theta}_{i,j}^{\dagger},\hat{\theta}_{i,j}^{\dagger}\right\} =\left\{\hat{\theta}_{i,j},\hat{\theta}_{i,j}\right\}=0$.
Thus, we can define non-local Majorana qubits as $\hat{Q}_{i,j}=2\hat{\theta}_{i,j}^{\dagger}\hat{\theta}_{i,j}-1$, which have eigenvalues $\pm 1$. Expressing $\hat{Q}_{i,j}(t)=\frac{1}{2}\left ( \hat{\gamma}_{2i-1}(t)-i\,\hat{\gamma}_{2j}(t) \right )\left ( \hat{\gamma}_{2i-1}(t)+i\,\hat{\gamma}_{2j}(t) \right )-1=\frac{1}{2}\left [ \hat{c}_{i}(t)+\hat{c}_{i}^{\dagger}(t), \hat{c}_{j}(t)-\hat{c}_{j}^{\dagger}(t)\right ]$ in terms of commutators of diagonal fermionic mode operators, it becomes possible to evaluate the corresponding TTC not only for the ground state but for any excited eigenstate $\ket{\psi_{K}}$ of the Kitaev chain. The TTC for general Majorana qubits  turns out to be
\begingroup\makeatletter\def\f@size{9.1}\check@mathfonts
\def\maketag@@@#1{\hbox{\m@th\large\normalfont#1}}
\begin{widetext}
\begin{equation}\label{cj3}
\begin{split}
\mathcal{C}_{i,j}\pap{t}=1-\sum_{k=1}^{N}\sum_{q=1}^{N}\Biggl[&\sin^2\pap{\frac{\eps_{k}+\eps_{q}}{2}t}\left [ \left ( u_{2k,i}+v_{2k,i}\right )\left ( u_{2q,j}-v_{2q,j}\right )-\left ( u_{2q,i}+v_{2q,i}\right )\left ( u_{2k,j}-v_{2k,j}\right )\right ]^2\left [ 1-\pap{n_{q}-n_{k}}^2\right ]\\
&+\sin^2\pap{\frac{\eps_{k}-\eps_{q}}{2}t}\left [ \left ( u_{2k,i}+v_{2k,i}\right )\left ( u_{2q,j}-v_{2q,j}\right )+\left ( u_{2q,i}+v_{2q,i}\right )\left ( u_{2k,j}-v_{2k,j}\right )\right ]^2\pap{n_{q}-n_{k}}^2\Biggr].
\end{split}
\end{equation}
\end{widetext}
\endgroup
where $n_k=0$ denotes the k-th fermion mode is empty while $n_k=1$ means it is occupied. By focusing on the edge TTC, i.e. $i=j=1$ and $\mathcal{C}_{1,1}\pap{t}=\mathcal{C}_{1}^{(z)}\pap{t}$, expanding the right hand side of Eq.~\eqref{cj3} up to second order in time and comparing it to the universal result quoted in Eq.~\eqref{otoc3}, a new identity results as
\begin{equation}\label{otoc10}
\sum_{k=1}^{N}\sum_{q=1}^{N}\left ( \eps_{k}+\eps_{q}\right )^2(u_{2 k,1}v_{2 q,1}-u_{2 q,1}v_{2 k,1})^2=2\Delta^2
\end{equation}
 which is valid for both open boundary Kitaev and transverse field Ising models (with $\Delta$ replaced by the spin exchange interaction) for arbitrary chain lengths. As before the identity given by Eq.~\eqref{otoc10} turns out to be another important consistency check for numerical calculations.\\
\section{Results and discussion} \label{sec_5}
 Now we evaluate numerically the different time correlations discussed in Sec.~\ref{sec_4}. All the results we describe below correspond to an open-ended Majorana fermion chain with $N=101$ sites in the many-body ground-state, $\ket{\psi_{K}}=\bigotimes_{k=1}^{N}\ket{0}$, with symmetric hopping-pairing energies, i.e. $\omega=\Delta=1$, which also fixes the energy scale. Their inverse fixes the time scale through the dimensionless variable $\Delta t$. In Figure~\ref{fig_2} the time evolution of both single edge Majorana qubits $\mathcal{C}_{1}^{(x)}\pap{t}$ and two-Majorana edge qubits $\mathcal{C}_{1}^{(z)}\pap{t}$ and $\mathcal{C}_{1,N}\left(t\right)$  is displayed for three specific values of the chemical potential, namely $\mu/\Delta=1.5$ (left panel), $\mu/\Delta=2.0$ (central panel) and $\mu/\Delta=2.5$ (right panel). Oscillatory features are dominant for both short- and intermediate-time regimes $\Delta t < 10$, which subsequently are attenuated until the TTCs reach stationary or asymptotical values for $\Delta t > 10$. We discuss first this long-time regime.

\begin{figure}[b!]
  \includegraphics[width=0.4 \textwidth]{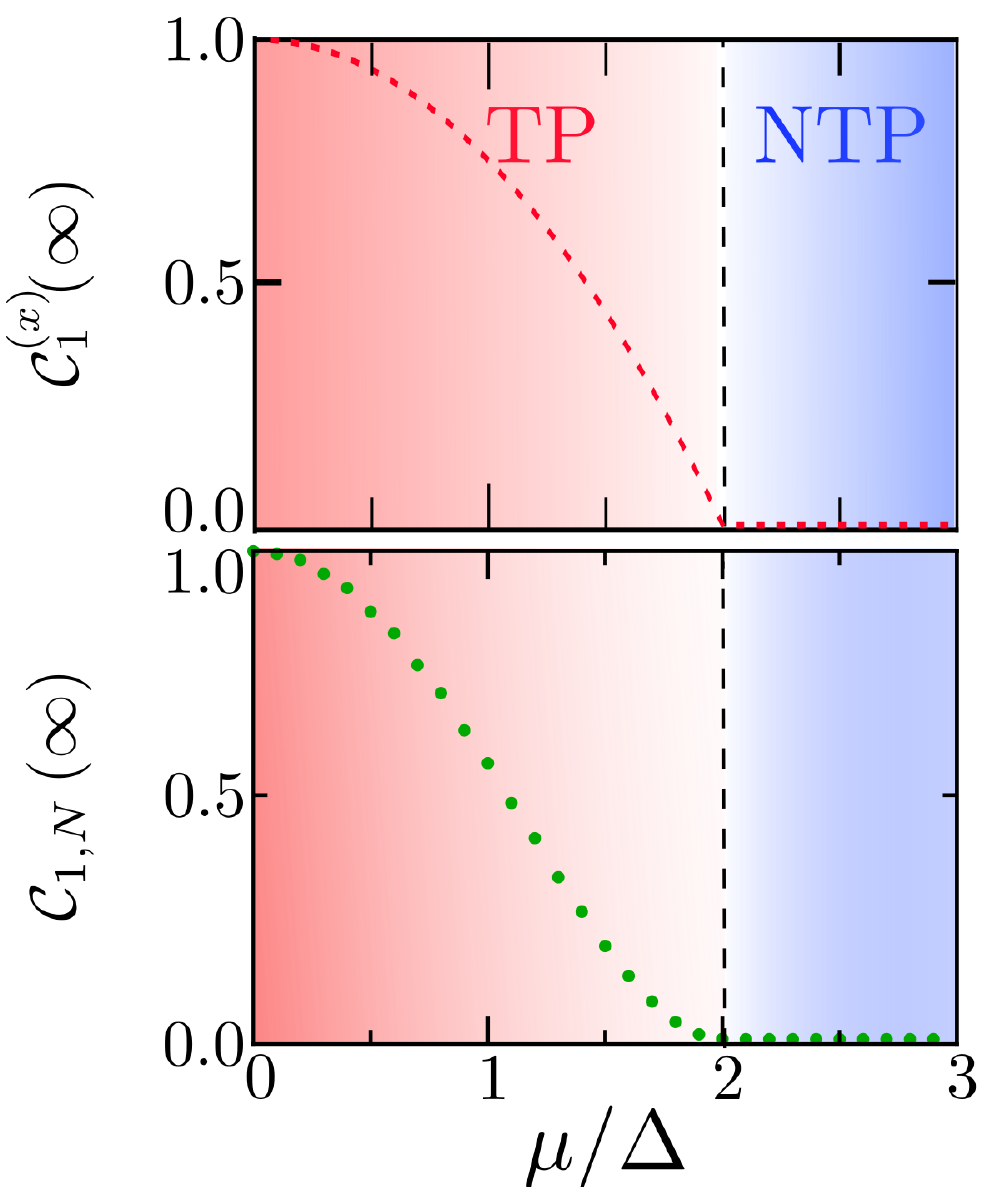}
  \caption{(color online)  Long-time limits of edge TTCs as a function of $\mu/\Delta$: top panel, single-Majorana edge $\mathcal{C}_{1}^{(x)}\left(t\right)$-TTC and bottom panel, non-local double-Majorana edge $\mathcal{C}_{1,N}\left(t\right)$-TTC. The order-parameter-like behavior exhibited by the long-time limits is evident. TP: topological phase, NTP: non-topological phase.}
  \label{fig_3}
\end{figure}
It can be seen that the asymptotic behaviors of $\mathcal{C}_{1}^{(x)}\pap{t}$ and $\mathcal{C}_{1,N}\left(t\right)$-TTCs are very different from that of $\mathcal{C}_{1}^{(z)}\pap{t}$-TTC crossing the critical point to the trivial phase. We found that these three TTCs remain finite in the topological phase even at infinity time, which agrees with the numerically-based observation in Ref.~\cite{Jack} that long coherence times for edge sites in open boundary Majorana fermion chains are possible. However, the long-time limits of $\mathcal{C}_{1}^{(x)}\pap{t}$ and $\mathcal{C}_{1,N}\left(t\right)$ vanish when the system enters the non-topological or trivial phase ($\mathcal{C}_{1}^{(z)}\pap{t}$ saturates to finite values at both phases). This order-parameter-like behavior of the TTC long-time limit is displayed in Figure~\ref{fig_3}. Furthermore, by both numerical fitting as well as the exact general duality property expressed in Eq.~\eqref{naraya3}, we establish that the long-time limit of the single-Majorana edge $\mathcal{C}_{1}^{(x)}(t)$-TTC has a simple specific functional behavior given by:
\begin{equation}\label{orpar}
\lim_{t\to\infty}\mathcal{C}_{1}^{(x)}(t)=\begin{cases}
1-\left (\frac{\mu}{2\Delta}\right )^{2} & \text{for}\; \mu<2\Delta\\
0  &\text{for}\; \mu>2\Delta.
\end{cases}
\end{equation}
\begin{figure}[t!]
  \includegraphics[width=0.48 \textwidth]{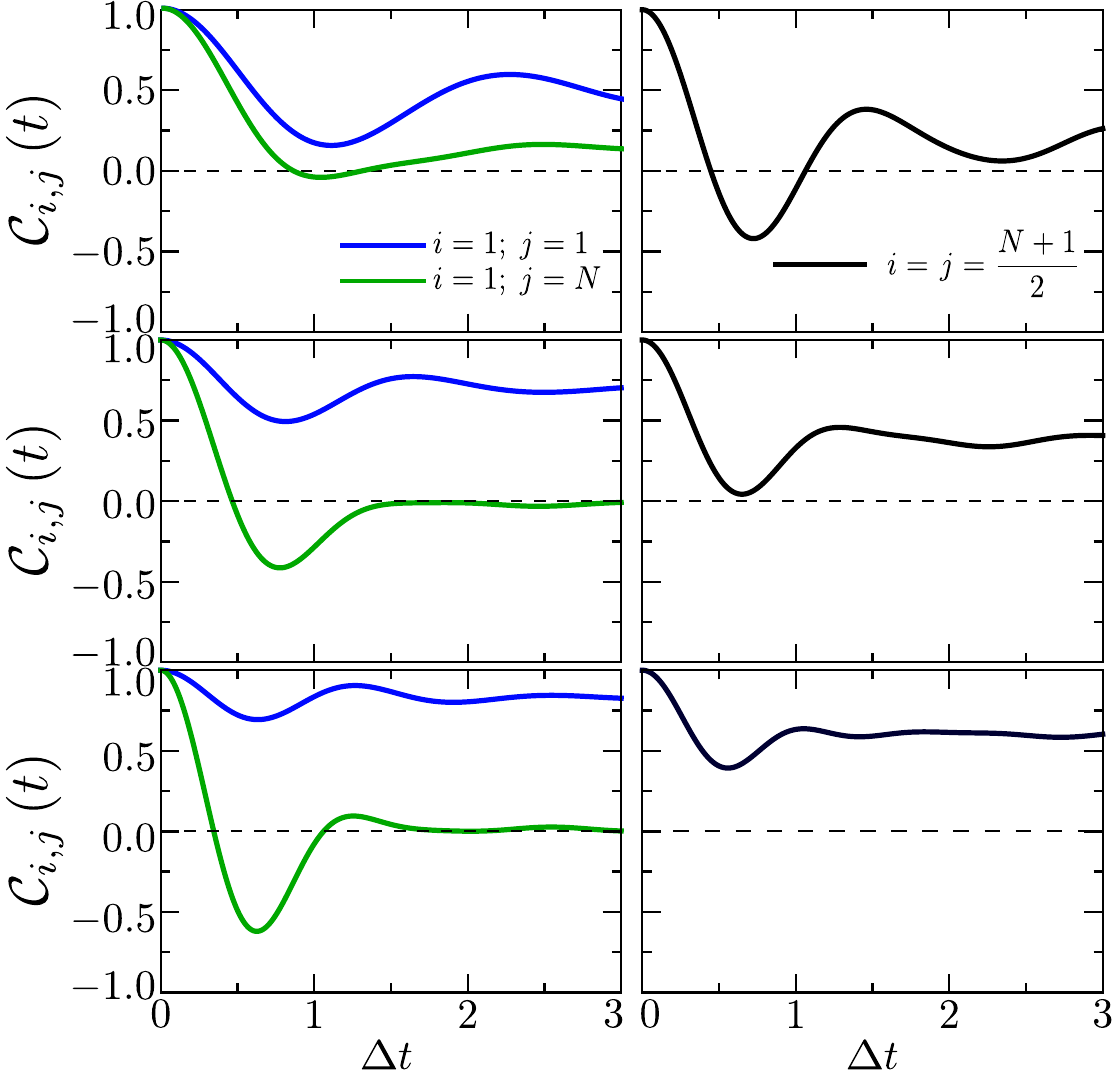}
  \caption{(color online) $\mathcal{C}_{i,j}\left(t\right)$-TTC as a function of the dimensionless time $\Delta t$ in the short- and intermediate-time regimes. Left panels display the TTC for {\em two-Majorana qubits}: blue (green) line $\mathcal{C}_{1,1}\pap{t}=\mathcal{C}_{1}^{(z)}\pap{t}$ local TTC ($\mathcal{C}_{1,N}\pap{t}$ non-local TTC), respectively. Right panels illustrate the time evolution behavior of TTC for a local {\em bulk} two-Majorana qubit (middle site of the Majorana fermion chain $\mathcal{C}_{\frac{N+1}{2},\frac{N+1}{2}}\pap{t}$). The chemical potentials are $\mu/\Delta=1.5$ (upper panels), $\mu/\Delta=2.0$ (middle panels), and $\mu/\Delta=2.5$ (bottom panels). }
  \label{fig_4}
\end{figure}
On the other hand, the decay of the limit value of the non-local $\mathcal{C}_{1,N}\left(t\right)$-TTC as a function of $\mu/\Delta$ has been evaluated numerically, showing a gradual transition, instead of an abrupt one, from one phase to the other. Note that these results are strictly valid for an infinitely long chain or for times below a certain limit where finite size effects could emerge, such as possible interference or revivals coming from the reflected influence of the other edge (not shown here). In addition, the quantum behavior of single-site TTC for the edge single- and double- Majorana qubits is similar to the $x$ and $z$ spin correlations of the transverse Ising model, and consequently its quantum critical point could also be detected by TTC measurements~\cite{gomez-ruiz}.\\
\\
Finally, we end this sub-section with a comparison between edge vs. bulk TTCs. In Figure~\ref{fig_4} the short- and intermediate-time behaviors of $\mathcal{C}_{i,j}\left(t\right)$-TTC are illustrated for edge-Majorana qubits, namely the local case ${i,j}={1}$ and the non-local case ${i,j}={1,N}$, and a bulk-two-Majorana qubit ${i,j}={\frac{N+1}{2},\frac{N+1}{2}}$. We conclude that apart from a different oscillation amplitude, the local two-Majorana TTCs, either located at the edge or at a bulk site, are very similar in going to a finite long-time limit in any phase, thus not being able to detect such phase transition by looking at that specific feature. This behavior contrasts with the one offered by the two-Majorana non-local edge TTC or even, as discussed above, with that shown by the single-Majorana edge TTC. Next, we focus on the consequences of these TTCs features when assessing macroscopic quantum coherence through the Leggett-Garg inequality violations, by both local- and non-local-TTCs.

\section{Leggett-Garg inequality}\label{sec_6}
Leggett and Garg~\cite{leggett,Leggett2} showed that temporal correlations obey similar inequalities as spatial non-local measurements such as those performed in a Bell inequality test set up. They approached this by first codifying our intuition about the macroscopic world into three principles: 
\begin{enumerate}[(i)]
\item Macroscopic realism: a system's property is well defined at every time regardless of being observed or not.

\item  Non-invasive measurability:  the system's evolution is unaffected by measurements taken on it.

\item Arrow of time: the outcome of a measurement cannot be affected by a subsequent measurement.
\end{enumerate}
We will focus on the following form of a LGI,
\begin{equation}
\mathcal{C}_{i,j}\left(t_{2}-t_{1}\right)+\mathcal{C}_{i,j}\left(t_{3}-t_{2}\right)-\mathcal{C}_{i,j}\left(t_{3}-t_{1}\right)\leq 1, \label{Eq:LGI1}
\end{equation}
where $\mathcal{C}_{i,j}\left(t_{\alpha},t_{\beta}\right )$ is a two-time correlation (see Eq.~\eqref{eq:Correl}) of the qubit nonlocal Majorana operator $\hat{Q}_{i,j}$ (with eigenvalues $\pm1$) between times $t_{\alpha}$ and  $t_{\beta}$, and $t_1 < t_2 < t_3$. We concentrate in the case of identical time intervals, i.e. $t_2-t_1=t_3-t_2=t$, defining a LGI function $\mathcal{K}_{i,j}(t)$ such as \cite{gomez-ruiz}:
\begin{equation}\label{lgi}
\mathcal{K}_{i,j}\pap{t}=2\mathcal{C}_{i,j}\left(t\right)-\mathcal{C}_{i,j}\left(2t\right) \leq 1.
\end{equation}
\begin{figure}[b!]
  \includegraphics[width=0.48 \textwidth]{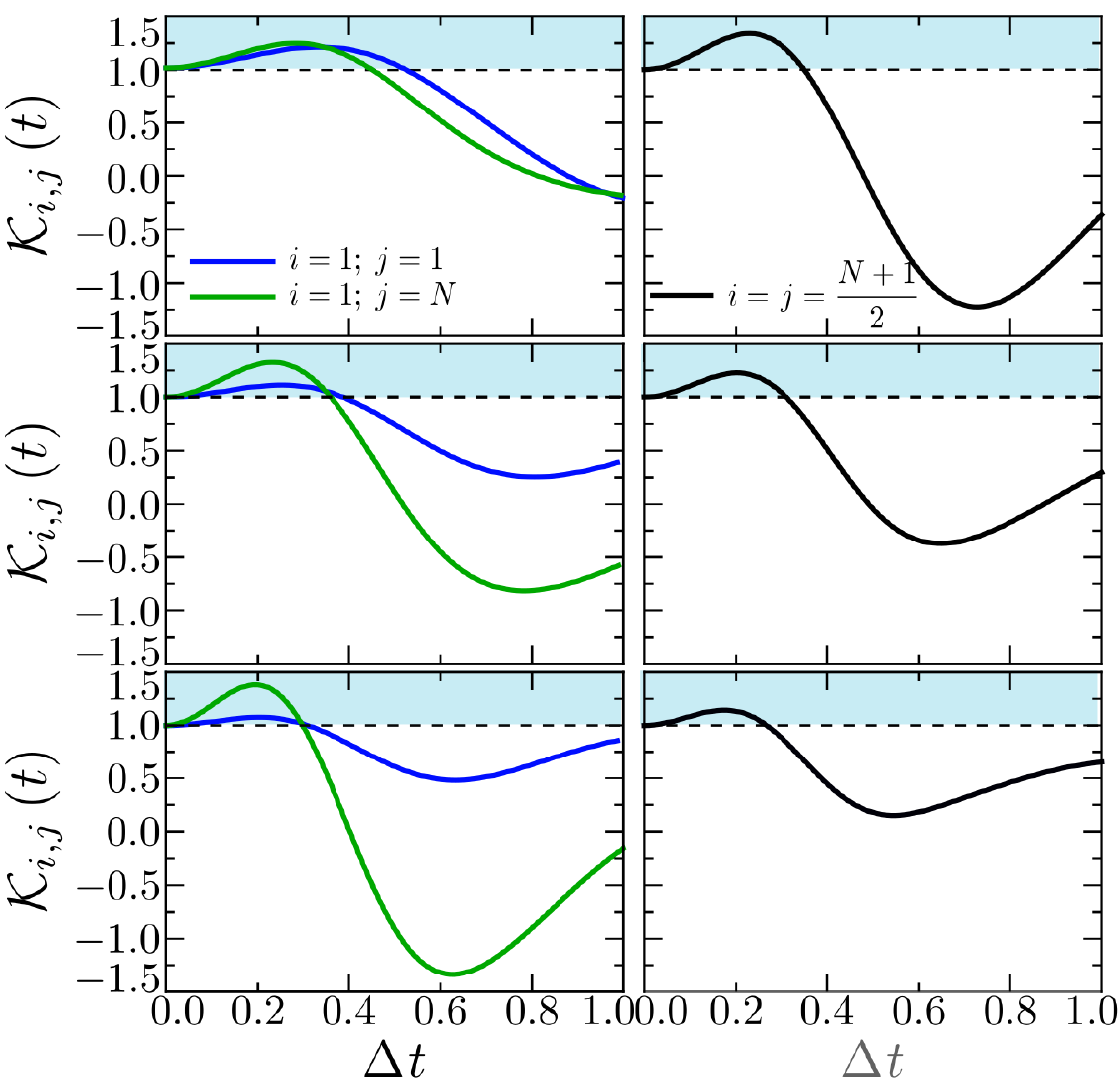}
  \caption{(color online) Two-Majorana $\mathcal{K}_{i,j}\pap{t}$ LGI function as a function of $\Delta t$. Panels and color lines have the same meaning as in Figure~\ref{fig_4}. The upper blue zones represent violations of the LGI given by Eq.~\eqref{lgi}.}
  \label{fig_5}
\end{figure}
Similarly to the Bell inequality test, any system that violates this LGI can be assured to behave in a nonclassical sense. From now on, we will take larger violations to LGI as an indication that a system has more quantum characteristics than another one.
\begin{figure}[t!]
  \includegraphics[width=0.5 \textwidth]{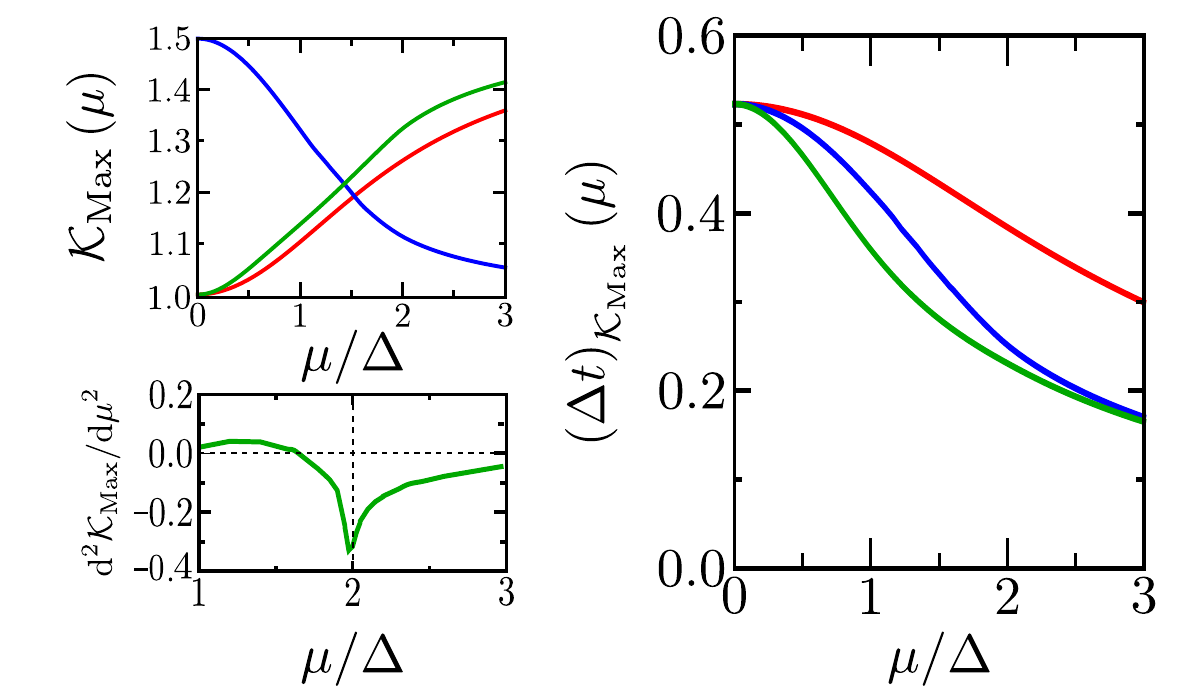}
  \caption{(color online) Left top panel: maximum violation of LGI as a function of $\mu/\Delta$. Right panel: time of maximum LGI violation as a function of $\mu/\Delta$. Left bottom panel: second derivative of the maximum LGI violation with respect to $\mu$ showing a dip signaling the phase transition for the non-local edges two-Majorana case. Red lines depict the $\mathcal{C}_{1}^{(x)}\pap{t}$-TTC based LGI, the blue lines represent the $\mathcal{C}_{1}^{(z)}\pap{t}$-TTC based LGI while the green lines correspond to $\mathcal{C}_{1,N}\pap{t}$-TTC based LGI.}
  \label{fig_6}
\end{figure}

Figure~\ref{fig_5} displays the evolution, as a function of $\Delta t$, of the LGI function $\mathcal{K}_{i,j}\pap{t}$ given by Eq.~\eqref{lgi} for the same parameters as used in Figure~\ref{fig_4}. We first note that the inequality is always violated at very early times, a result that can be already understood from the $\mathcal{O}(t^2)$ expansions given in Eqs.~\eqref{otoc2} and~\eqref{otoc3}. Specifically, the $\mathcal{C}_{1}^{(x)}\pap{t}$-TTC based LGI, denoted by $\mathcal{K}_1^{(x)}\pap{t}$, is given by
\begin{equation}
\mathcal{K}_1^{(x)}\pap{t}\simeq 1+\mu^2t^2+{\cal O}(t^4),
\end{equation}
while that based on $\mathcal{C}_{1}^{(z)}\pap{t}$, denoted by $\mathcal{K}_1^{(z)}\pap{t}$, is
\begin{equation}
\mathcal{K}_1^{(z)}\pap{t}\simeq 1+4\Delta^2t^2+{\cal O}(t^4).
\end{equation}
Thus, the initial growth of both inequality violations is captured again by the universal initial curvatures of the corresponding TTCs. Furthermore, the early-time violations for $\mathcal{K}_1^{(x)}\pap{t}$ and $\mathcal{K}_1^{(z)}\pap{t}$ become identical at $\mu=2\Delta$, i.e. the critical point. This conclusion provides an alternative route to identifying the topological phase transition.\\
\\
Now we consider different inequalities for longer times. It is evident that LGI functions based on two-local-Majorana TTCs such as edge $\mathcal{C}_{1,1}\pap{t}$ and bulk $\mathcal{C}_{\frac{N+1}{2},\frac{N+1}{2}}\pap{t}$ follow a similar trend, which is very different to that of the non-local two-Majorana TTC given by $\mathcal{C}_{1,N}\pap{t}$ when crossing from one phase to the other. The local LGI violations turn out to be stronger in the topological phase while the non-local LGI violation increases when passing from the topological to the trivial phase. This contrasting behavior can also be seen in Figure~\ref{fig_6}, where we compare the maximum LGI violation $\mathcal{K}_{\text{Max}}\pap{\mu}$ as a function of $\mu$ (left panel) for single- and double-Majorana qubits, as well as the times for which that maximum violation occurs $t_{\mathcal{K}_{\text{Max}}}\pap{\mu}$ for the same qubits (right panel).  Interestingly, for the non-local edges two-Majorana case, the second derivative of $\mathcal{K}_{\text{Max}}\pap{\mu}$ with respect to $\mu$ shows a dip signaling the phase transition, again an inherited feature from the corresponding time correlations $\mathcal{C}_{1,N}\pap{t}$ (see Fig.~\ref{fig_1}). Thus, we can conclude that LGI violations by non-local Majorana qubits are sensitive to the topological features of the underlying phase, and consequently they could be explored in properly designed experimental setups.

\section{Experimental implementations} \label{sec_7}
Among the most promising candidates for experimentally detecting Majorana edge fermions in condensed matter systems are chains of magnetic atoms on superconducting surfaces~\cite{Nadj,HaltermanPRB,JianPRB} and semiconducting nanowires with large Rashba spin-orbit interaction under an applied magnetic field and induced superconductivity by proximity effects~\cite{SticletPRL,Rashba1}. Previous works have focused on local sensitive tunneling signatures of the topological phase transition in the boundary fermion occupation (Kitaev chain) or boundary spin (transverse field Ising chain).

In the Rashba nanowire setup Sticlet et al.~\cite{SticletPRL} define local Majorana pseudo-spins and argue that they could be measured by spin-polarized STM allowing to directly visualize the Majorana fermionic states and to test the topological character of the 1D system. On the other hand, Deng et al.~\cite{DengArxiv} reported that highly sensitive experiments have been recently conducted where the non-locality of Majorana qubits can be {\it locally} probed by a quantum dot at one end of the nanowire. These state-of-the-art experiments could evolve to develop time dependent sensitivity as required for detecting local and nonlocal TTCs. Recently, there has been great interest in contrasting distinctive signatures of spin polarization for Andreev and Majorana bound states~\cite{ZhangPRB} since when identifying topological phases effects coming from the presence of quasiparticle states inside the superconducting gap should be carefully eliminated~\cite{ChienPRB}. Thus, it is most desirable to have additional signatures available (besides tunneling conductance signatures of Majorana fermions) that would allow one to identify the topological phase transition. It has been proposed in~\cite{SticletPRL} and~\cite{DengArxiv} how to distinguish such differences between Andreev vs. Majorana signatures by accessing true nonlocal features. In this way, our results as given by the behavior of local $\mathcal{C}_{1}^{(x)}\left(t\right)$ and most importantly by nonlocal $\mathcal{C}_{1,N}\left(t\right)$, and their LGI combinations, should be relevant for extending that kind of search of true Majorana behavior.

Furthermore, recently, spin noise spectroscopy has been shown as a powerful tool to experimentally accessing the autocorrelation function~\cite{NikolaiRPP,BurnellPRB2}. The universal short-time behavior described by Eqs.~\eqref{otoc1} and~\eqref{otoc2} could be exploited in spin fluctuation measurements as an alternative route to get information about the dynamics~\cite{BurnellPRB}. Such rich variety of behaviors would also permit the study of temporal effects as well as different kind of susceptibilities, through their Fourier transform equivalents, in topological quantum computing settings.

Therefore, in light of recent experiments, we demonstrate in the present work that TTC and LGI behaviors exhibit a quantum-phase sensitive signature due to the appearance of zero-energy-modes in the topological phase that will manifest themselves in the long-time behavior of both local as well as nonlocal qubit TTCs. This provides an experimentally useful diagnostic tool to detect topological phase transitions.

\section{Concluding remarks} \label{sec_8}
In summary, we have provided evidence that time correlations and violations of LGI establish new testable signatures of topological phase transitions. The behavior of that sort of inequality is a direct consequence of time correlations in local and nonlocal Majorana qubits. Specifically, we have identified signatures of the MFC topological phase transition in any of three time domains: (i) in the short-time limit we found universal features such as the out-of-time-ordered correlation and a dip in the second $\mu$-derivative marking the phase transition; (ii) in the intermediate time region, the LGI violations are sensitive to the quantum phase the system is; and (iii) in the long-time limit, the asymptotic values of single- and double-Majorana edge TTCs act as order-parameter-like indicators. Specifically, we propose that a measurement of the long time saturation value of the local edge $\mathcal{C}_{1}^{(x)}$-TTC as well as the non-local edge $\mathcal{C}_{1,N}$-TTC provide a witness of the topological ($\neq 0$) vs. non-topological ($=0$) phase transition of Majorana fermion chain systems, as it probes directly the existence of zero energy modes. Additionally, in the former case it gives direct access to the electron-hole weight of such modes. The results are especially relevant because the whole question of quantum coherence in complex mesoscopic systems is taking up a new impulse in the community and is of interest to researchers not only in quantum information and foundations but also in condensed matter.
\begin{acknowledgments}
FJG-R, JJM-A, FJR and LQ acknowledge financial support from Vice-Rectoría Investigaciones through UniAndes-2015 project \emph{Quantum control of nonequilibrium hybrid systems-Part II}. F.J.G.R and F.J.R acknowledges financial support from Facultad de Ciencias at Universidad de los Andes (2018-I). CT acknowledges support from the spanish MINECO under contrats MAT2014-53119-C2-1-R and MAT2017-83722-R. We acknowledge financial support from the project CEAL-AL/2017-25 UAM-Banco Santander for a collaboration between the Universidad Autónoma de Madrid and the Universidad de los Andes at Bogota. LQ thanks UAM for kind hospitality.  FJG-R thanks UMass Boston for kind hospitality.
\end{acknowledgments}
\bibliography{MyBib_Majorana}	

\begin{thebibliography}{58}
\bibitem [{\citenamefont {url will be inserted~by publisher}()}]{SM_Gomez}%
  \BibitemOpen
  \bibfield  {author} {\bibinfo {author} {\bibnamefont {url will be inserted~by
  publisher}},\ }\href@noop {} {\ }\BibitemShut {NoStop}%
\end{thebibliography}
\newpage
\pagebreak
\clearpage
\widetext
\begin{center}
\textbf{\large Supplemental Material: Universal two-time correlations, out-of-time-ordered correlators and Leggett-Garg inequality violation by edge Majorana fermion qubits}\\
\vspace{0.5cm}
F. J. Gómez-Ruiz$^{1,2}$, J. J. Mendoza-Arenas$^{1}$, F. J. Rodríguez$^1$, C. Tejedor$^3$, and L. Quiroga$^1$\\
$^1${\it Departamento de Física, Universidad de los Andes, A.A. 4976, Bogotá D. C., Colombia.}\\
$^2${\it Department of Physics, University of Massachusetts, Boston, MA 02125, USA}\\
$^3${\it Departamento de Física Teórica de la Materia Condensada and Condensed Matter Physics Center (IFIMAC), Universidad Autónoma de Madrid, 28049, Spain.}
\end{center}

\begin{center}
\begin{tabular}{p{14cm}}
\vspace{0.1cm}
\quad In this Supplementary Material (SM), we provide details of the analytical strategies employed in the main text to obtain exact numerical results for two-time correlations, out-of-time-ordered correlators and Leggett-Garg inequalities for edge Majorana fermion qubits.      
\end{tabular}
\end{center}
\setcounter{equation}{0}
\setcounter{figure}{0}
\setcounter{table}{0}
\setcounter{section}{0}
\setcounter{page}{1}
\makeatletter
\renewcommand{\theequation}{S\arabic{equation}}
\renewcommand{\thefigure}{S\arabic{figure}}
\renewcommand{\bibnumfmt}[1]{[S#1]}
\renewcommand{\citenumfont}[1]{S#1}

\section{Exact diagonalization of the Kitaev Hamiltonian: Bogoliubov-de Gennes approach}
Since the Kitaev Hamiltonian is quadratic in fermionic operators $\hat{c}_{j}$ and $\hat{c}_{j}^{\dagger}$, its exact diagonalization via a Bogoliubov-de Gennes transformation is always feasible~\cite{Olesia}. The matrix representation of the Kitaev Hamiltonian, see Eq.(1) of the main text, takes the form:
\begin{equation} \label{Matrix_Kitaev}
\hat{H}=\frac{1}{2}\begin{pmatrix}
\hat{c}_1^{\dagger} & \hat{c}_{1} & \cdots & \hat{c}_N^{\dagger} & \hat{c}_{N}
\end{pmatrix}\left(\begin{array}{cccccccccc}
-\mu & 0 & -\Delta & -w & 0 & 0 & 0 & \cdots  &0& 0\\
0 & \mu & w & \Delta & 0 & 0 & 0 &\cdots  & 0 & 0\\
-\Delta & w & -\mu & 0 & -\Delta & -w & 0 &  \cdots & 0 & 0\\
-w &\Delta & 0 &\mu &w & \Delta &0 &\cdots & 0 & 0\\
0 & 0 &-\Delta & w & \ddots & \ddots & \ddots&\ddots  &\vdots &\vdots \\
0 & 0 & -w & \Delta & \ddots & \ddots & \ddots & \ddots  & \ddots&\vdots \\
0 & 0 & 0 & 0 & \ddots  &\ddots & \ddots & \ddots &   \ddots & \vdots\\
\vdots&\vdots&\vdots&\vdots&\vdots&\vdots&\vdots&\vdots&\vdots&\vdots\\
0&0&0&0&\cdots&0&-\Delta&w&-\mu&0\\
0&0&0&0&\cdots&0&-w&\Delta&0&\mu
\end{array}
\right)\begin{pmatrix}
\hat{c}_1\\
\hat{c}_{1}^{\dagger}\\
\vdots\\
\vdots\\
\vdots\\
\vdots\\
\vdots\\
\hat{c}_N\\
\hat{c}_{N}^{\dagger}
\end{pmatrix}.
\end{equation}
The central $2N\times 2N$ matrix in Eq.~\eqref{Matrix_Kitaev}, which we denote by $\hat{\mathcal{H}}$, can be rendered to a diagonal form $\hat{\mathcal{H}}_D=\hat{D}^{-1}\hat{\mathcal{H}}\hat{D}$ by a unitary matrix such as:
\begin{equation}\label{duv1}
\hat{D}=\left(\begin{array}{cccccccccc}
u_{1,1} & u_{2,1} & u_{3,1} & u_{4,1} & \cdots & \cdots & \cdots & \cdots  &u_{2N-1,1}& u_{2N,1}\\
v_{1,1} & v_{2,1} & v_{3,1} & v_{4,1} & \cdots & \cdots & \cdots & \cdots  &v_{2N-1,1}& v_{2N,1}\\
u_{1,2} & u_{2,2} & u_{3,2} & u_{4,2} & \cdots & \cdots & \cdots & \cdots  &u_{2N-1,2}& u_{2N,2}\\
v_{1,2} & v_{2,2} & v_{3,2} & v_{4,2} & \cdots & \cdots & \cdots& \cdots  &v_{2N-1,2}& v_{2N,2}\\
\vdots & \vdots &\vdots & \vdots & \ddots & \ddots & \ddots&\ddots  &\vdots &\vdots \\
\vdots & \vdots &\vdots & \vdots & \ddots & \ddots & \ddots&\ddots  &\vdots &\vdots \\
u_{1,N} & u_{2,N} & u_{3,N} & u_{4,N} & \cdots & \cdots & \cdots & \cdots  &u_{2N-1,N}& u_{2N,N}\\
v_{1,N} & v_{2,N} & v_{3,N} & v_{4,N} & \cdots & \cdots & \cdots & \cdots  &v_{2N-1,N}& v_{2N,N}
\end{array}
\right).
\end{equation}
Since $u_{2q-1,j}=v_{2q,j}$ and $v_{2q-1,j}=u_{2q,j}$, the unitary property of matrix $\hat{D}$ implies that
\begin{align}\label{diag1}
\sum_{q=1}^{N}\pas{u_{2q,i}u_{2q,j}+v_{2q,i}v_{2q,j}}=\delta_{i,j},\qquad\qquad\sum_{q=1}^{N}u_{2q,i}v_{2q,j}=0
\end{align}
for every site $j=1,...,N$. The diagonal matrix $\hat{\mathcal{H}}_D$ is ordered as
\begin{equation}
\hat{\mathcal{H}}_D=\left(\begin{array}{cccccccccc}
-\epsilon_1 & 0 & \cdots & 0 & 0\\
0 & \epsilon_1 & \cdots & 0 & 0\\
\vdots & \vdots & \ddots  &\vdots & \vdots\\
0 & 0 & \cdots & -\epsilon_N & 0\\
0 & 0 & \cdots & 0 & \epsilon_N
\end{array}
\right)
\end{equation}
where, for $\omega=\Delta$, the positive energies are given by:
\begin{equation}\label{diag2}
\epsilon_k=-\mu\sum_{j=1}^N\left [ u_{2k,j}^2-v_{2k,j}^2 \right ]-2\Delta \sum_{j=1}^{N-1}\left [ u_{2k,j}-v_{2k,j} \right ]\left [ u_{2k,j+1}+v_{2k,j+1} \right ].
\end{equation}
From the entries of matrix $\hat{D}$ in Eq.~\eqref{duv1} the standard  Bogoliubov-de Gennes transformation to new annihilation (creation) fermionic operators $\hat{d} \pap{\hat{d}^{\dagger}}$ can be written as:
\begin{align}\label{Bogol_tra}
\hat{c}_{j} &= \sum_{k=1}^{N} \left ( u_{2k,j} \hat{d}_{k} + v_{2k,j}\hat{d}^{\dagger}_{k} \right ), \qquad\qquad\hat{c}_{j}^{\dagger} = \sum_{k=1}^{N} \left ( u_{2k,j} \hat{d}_{k}^{\dagger} + v_{2k,j}\hat{d}_{k} \right ).
\end{align}
It can be easily checked that canonical fermionic anticommutation relations for local operators $\hat{c}_{j}$ imply that the same relations hold for the fermionic mode operators $\hat{d}_{k}$, that is
	\[\left \{\hat{d}_{k},\hat{d}_{k'}^{\dagger}\right \}=\delta_{k,k'},\qquad\left\{\hat{d}_{k}^{\dagger},\hat{d}_{k'}^{\dagger}\right\} =\left\{\hat{d}_{k},\hat{d}_{k'}\right\}=0.
\]
\section{Exact diagonalization of the Kitaev Hamiltonian: Majorana approach}
\begin{figure}[h!]
\begin{center}
\includegraphics[scale=0.7]{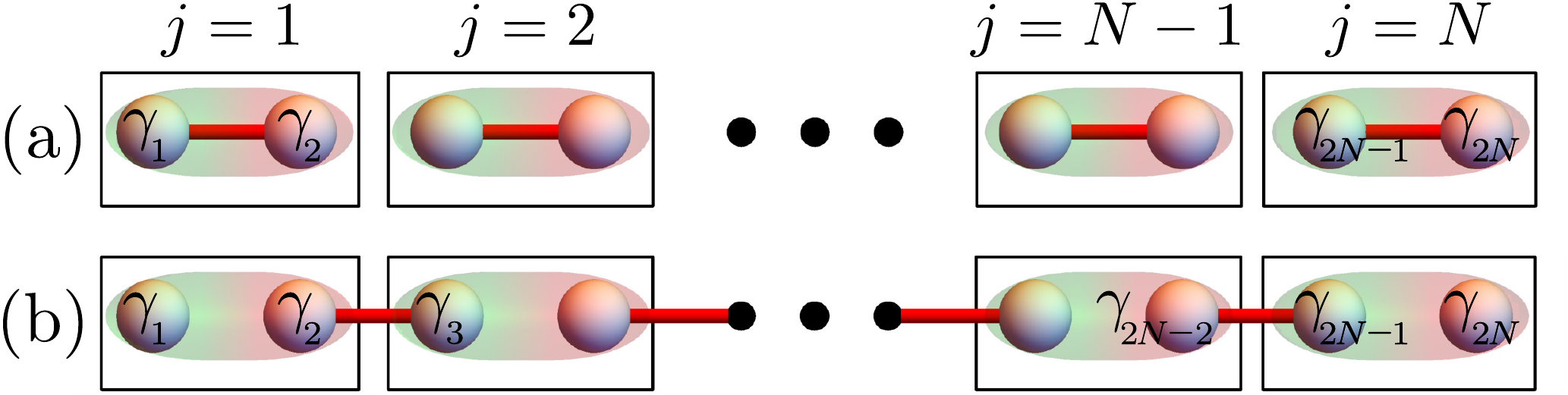}
\caption{\label{Apar_1} Schematic illustration of the Kitaev-Majorana Hamiltonian, {\bf (a)} in the trivial limit, and {\bf (b)} in the nontrivial limit with coupling between Majorana fermions $\hat{\gamma}_{2j}$ and $\hat{\gamma}_{2j+1}$ only. The solid spheres represent the Majorana fermions $\gamma_{2j-1}$ and $\gamma_{2j}$ making up each physical $j$ site in the Kitaev chain. In the nontrivial phase, the zero energy Majorana Modes (ZEM) are present at the left and right boundaries of the lattice, which are illustrated by the two unpaired spheres.}
\end{center}
\end{figure}

To diagonalize the Kitaev model~\cite{kitaev2} within the Majorana approach we proceed as follows. We formally define {\it Majorana operators} as a combination of the creation and annihilation fermionic operators, namely
\begin{align}\label{MaO}
\hat{\gamma}_{2j-1}=\hat{c}_j + \hat{c}_j^{\dagger}, \qquad& \qquad\hat{\gamma}_{2j}=-\ii\left(\hat{c}_j - \hat{c}_j^{\dagger}\right),
\end{align}
with $j=1,\ldots, N$. Therefore
\begin{align}\label{c-oper}
\hat{c}_j^{\dagger}=\frac{1}{2}\left ( \hat{\gamma}_{2j-1}-i\hat{\gamma}_{2j} \right ), \qquad& \qquad \hat{c}_j=\frac{1}{2}\left ( \hat{\gamma}_{2j-1}+i\hat{\gamma}_{2j} \right ).
\end{align}
By direct substitution of Eq.~\eqref{c-oper} into the  Kitaev Hamiltonian of Eq.(1) main text, we obtain the Kitaev-Majorana (KM) Hamiltonian,
\begin{equation}\label{KMH}
\hat{H}_{\text{KM}}=-\ii\frac{\mu}{2}\sum_{j=1}^{N}\hat{\gamma}_{2j-1}\hat{\gamma}_{2j}+\frac{\ii}{2}\sum_{j=1}^{N-1}\left[\left(\Delta+w\right)\hat{\gamma}_{2j}\gamma_{2j+1}-\left(w-\Delta\right)\hat{\gamma}_{2j-1}\hat{\gamma}_{2j+2}\right].
\end{equation}
The parameters $\mu$, $\Delta$ and $\omega$ induce relative complex interactions between the Majorana modes. Now we briefly explain two limit cases of the KM Hamiltonian.
\\{\em First limit case:} We start with the simplest case $\Delta=\omega=0$ with $\mu<0$, yielding to a Kitaev state in the so-called topologically trivial phase. Therefore, the KM Hamiltonian of Eq.~\eqref{KMH} takes the trivial form
\begin{align}\label{Hmajo}
\hat{H}_{\text{Trivial}}=-\ii\frac{\mu}{2}\sum_{j=1}^{N}\hat{\gamma}_{2j-1}\hat{\gamma}_{2j}.
\end{align}
Here only the first term of equation \eqref{KMH} is different from zero, leaving a coupling only between Majorana modes $\hat{\gamma}_{2j-1}$ and $\hat{\gamma}_{2j}$ at the same lattice site $j$, as Fig.~\ref{Apar_1}(a) schematically illustrates. This leads to a ground state with all occupation numbers equal to $0$.
\\{\em Second limit case:} We now consider the Hamiltonian of Eq.(1) with $\mu=0$, namely
\begin{equation*}
\hat{H}=\frac{\ii}{2}\sum_{j=1}^{N-1}\left[\left(\Delta+\omega\right)\hat{\gamma}_{2j}\gamma_{2j+1}-\left(\omega-\Delta\right)\hat{\gamma}_{2j-1}\hat{\gamma}_{2j+2}\right].
\end{equation*}
This last form simplifies even more when $\Delta=\omega>0$ to the compact expression $\hat{H}=\Delta\ii\sum_{j=1}^{N-1}\hat{\gamma}_{2j}\hat{\gamma}_{2j+1}$, indicating that Majorana operators from neighboring sites are paired together, so that the even numbered $\hat{\gamma}$ at site $j$ is coupled to the odd numbered $\hat{\gamma}$ at site $j+1$, as depicted in Fig.~\ref{Apar_1}(b). The first and last Majorana fermions are thus left unpaired, corresponding to the zero energy Majorana Modes (ZEM)~\cite{Leijnse,aguado}.

\section{Two-time correlations of Majorana qubits}
For a better understanding of our results and their implications, we provide a detailed description of the analytical approach used to obtain the two-time correlations (TTC) for qubits formed by any pair of Majorana fermions such as $\hat{\gamma}_{2i-1}$ and $\hat{\gamma}_{2i}$. In the main text we defined the operators $\hat{\theta}_{i,j}$ and $\hat{\theta}_{i,j}^{\dagger}$ (see Eq.(20)), and focused our attention on the dichotomic operator $\hat{Q}_{i,j}=2\hat{\theta}_{i,j}^{\dagger}\hat{\theta}_{i,j}-1$ for non-local Majorana qubits, which is an observable with eigenvalues $\pm1$. By direct substitution of the standard Bogoliubov-de Gennes transformation (Eq.~\eqref{Bogol_tra}), we can rewrite the operator $\hat{Q}_{i,j}$ as
\begin{equation}\label{majsup2}
\begin{split}
\hat{Q}_{i,j}&=\frac{1}{2}\left ( \hat{\gamma}_{2i-1}-i\,\hat{\gamma}_{2j} \right )\left ( \hat{\gamma}_{2i-1}+i\,\hat{\gamma}_{2j} \right )-1=\frac{1}{2}\left [ \hat{c}_{i}+\hat{c}_{i}^{\dagger}, \hat{c}_{j}-\hat{c}_{j}^{\dagger}\right ]\\
\hat{Q}_{i,j}&=\frac{1}{2}\sum_{k=1}^{N}\sum_{q=1}^{N}\left ( u_{2k,i}+v_{2k,i} \right )\left ( v_{2q,j}-u_{2q,j} \right )\left [ \hat{d}_{k}+\hat{d}_{k}^{\dagger}, \hat{d}_{q}-\hat{d}_{q}^{\dagger}\right ].
\end{split}
\end{equation}
Similarly we evaluate its time-evolution $\hat{Q}_{i,j}(t)=\ee^{\ii \hat{H}t}\hat{Q}_{i,j}(0)\ee^{-\ii \hat{H}t}$, obtaining that
\begin{equation}\label{majsup3}
\hat{Q}_{i,j}(t)=\frac{1}{2}\sum_{k=1}^{N}\sum_{q=1}^{N}\left ( u_{2k,i}+v_{2k,i} \right )\left ( u_{2q,j}-v_{2q,j} \right )\left [ \ee^{-\ii \epsilon_kt}\hat{d}_{k}+\ee^{\ii \epsilon_kt}\hat{d}_{k}^{\dagger}, \ee^{-\ii \epsilon_qt}\hat{d}_{q}-\ee^{\ii \epsilon_qt}\hat{d}_{q}^{\dagger}\right ].
\end{equation}
Our particular interest is to analyze the behavior of the two-time correlations throughout the Kitaev chain parameters, especially close to the critical point. Therefore, we explicitly calculate the symmetric TTC $\mathcal{C}_{i,j}\pap{t}=\frac{1}{2}\langle\lbrace \hat{Q}_{i,j}\pap{t},\hat{Q}_{i,j}\pap{0}\rbrace\rangle$, where $\lbrace\bullet,\bullet\rbrace$ is the anticommutator. The TTC for the general non-local Majorana qubit operator $\hat{Q}_{i,j}$ is found to be
\begin{equation}\label{ter_0}
\begin{split}
\mathcal{C}_{i,j}\pap{t}=&\frac{1}{8}\sum_{k=1}^{N}\sum_{q=1}^{N}\sum_{k'=1}^{N}\sum_{q'=1}^{N}\left ( u_{2k,i}+v_{2k,i} \right )\left ( v_{2q,j}-u_{2q,j} \right )
\left ( u_{2k',i}+v_{2k',i} \right )\left ( v_{2q',j}-u_{2q',j} \right )\\
&\bigg\langle  \biggl( \hat{d}_{k}\hat{d}_{q}\ee^{-\ii \pap{\eps_k +\eps_q}t}-\hat{d}_{k}\hat{d}_{q}^{\dagger}\ee^{-\ii \pap{\eps_k -\eps_q}t}+
\hat{d}_{k}^{\dagger}\hat{d}_{q}\ee^{\ii \pap{\eps_k -\eps_q}t}-\hat{d}_{k}^{\dagger}\hat{d}_{q}^{\dagger}\ee^{\ii \pap{\eps_k +\eps_q}t}\\
&\quad\qquad-
\hat{d}_{q}\hat{d}_{k}\ee^{-\ii \pap{\eps_k +\eps_q}t}-\hat{d}_{q}\hat{d}_{k}^{\dagger}\ee^{\ii \pap{\eps_k -\eps_q}t}+
\hat{d}_{q}^{\dagger}\hat{d}_{k}\ee^{-\ii \pap{\eps_k -\eps_q}t}+\hat{d}_{q}^{\dagger}\hat{d}_{k}^{\dagger}\ee^{\ii \pap{\eps_k +\eps_q}t}\biggr )\\
&\quad\qquad\biggl( \hat{d}_{k'}\hat{d}_{q'}-\hat{d}_{k'}\hat{d}_{q'}^{\dagger}+\hat{d}_{k'}^{\dagger}\hat{d}_{q'}-\hat{d}_{k'}^{\dagger}\hat{d}_{q'}^{\dagger}
-\hat{d}_{q'}\hat{d}_{k'}-\hat{d}_{q'}\hat{d}_{k'}^{\dagger}+\hat{d}_{q'}^{\dagger}\hat{d}_{k'}+\hat{d}_{q'}^{\dagger}\hat{d}_{k'}^{\dagger}
\biggr )\\
&\quad\qquad+\biggl( \hat{d}_{k}\hat{d}_{q}-\hat{d}_{k}\hat{d}_{q}^{\dagger}+\hat{d}_{k}^{\dagger}\hat{d}_{q}-\hat{d}_{k}^{\dagger}\hat{d}_{q}^{\dagger}
-\hat{d}_{q}\hat{d}_{k}-\hat{d}_{q}\hat{d}_{k}^{\dagger}+\hat{d}_{q}^{\dagger}\hat{d}_{k}+\hat{d}_{q}^{\dagger}\hat{d}_{k}^{\dagger}
\biggr )\\
&\quad\qquad\biggl( \hat{d}_{k'}\hat{d}_{q'}\ee^{-\ii \pap{\eps_{k'} +\eps_{q'}}t}-\hat{d}_{k'}\hat{d}_{q'}^{\dagger}\ee^{-\ii \pap{\eps_{k'} -\eps_{q'}}t}+
\hat{d}_{k'}^{\dagger}\hat{d}_{q'}\ee^{\ii \pap{\eps_{k'} -\eps_{q'}}t}-\hat{d}_{k'}^{\dagger}\hat{d}_{q'}^{\dagger}\ee^{\ii \pap{\eps_{k'} +\eps_{q'}}t}\\
&\quad\qquad-
\hat{d}_{q'}\hat{d}_{k'}\ee^{-\ii \pap{\eps_{k'} +\eps_{q'}}t}-\hat{d}_{q'}\hat{d}_{k'}^{\dagger}\ee^{\ii \pap{\eps_{k'} -\eps_{q'}}t}+
\hat{d}_{q'}^{\dagger}\hat{d}_{k'}\ee^{-\ii \pap{\eps_{k'} -\eps_{q'}}t}+\hat{d}_{q'}^{\dagger}\hat{d}_{k'}^{\dagger}\ee^{\ii \pap{\eps_{k'} +\eps_{q'}}t}\biggr )
\bigg\rangle,
\end{split}
\end{equation}
 In order to proceed further from Eq.~\eqref{ter_0} we note that only elements with equal number of creation and annihilation operators are relevant since only they can produce nonvanishing expectation values for eigenstates of the system with a well defined number of elementary fermionic excitations. These nonvanishing terms turn out to be
\begin{equation}\label{nnt2}
\begin{split}
\big\langle\hat{d}_{k}\hat{d}_{q}^{\dagger}\hat{d}_{k'}\hat{d}_{q'}^{\dagger}\big\rangle&=\delta_{k,q}\delta_{k',q'}(1-n_{k})(1-n_{k'})+\delta_{k,q'}\delta_{k',q}(1-n_{k})n_{k'}\\
\big\langle\hat{d}_{k}\hat{d}_{q}^{\dagger}\hat{d}_{k'}^{\dagger}\hat{d}_{q'}\big\rangle&=\delta_{k,q}\delta_{k',q'}(1-n_{k})n_{k'}-\delta_{k,k'}\delta_{q,q'}(1-n_{k})n_{q}\\
\big\langle\hat{d}_{k}^{\dagger}\hat{d}_{q}\hat{d}_{k'}\hat{d}_{q'}^{\dagger}\big\rangle&=\delta_{k,q}\delta_{k',q'}n_{k}(1-n_{k'})-\delta_{k,k'}\delta_{q,q'}n_{k}(1-n_{q})\\
\big\langle\hat{d}_{k}^{\dagger}\hat{d}_{q}\hat{d}_{k'}^{\dagger}\hat{d}_{q'}\big\rangle&=\delta_{k,q}\delta_{k',q'}n_{k}n_{k'}+\delta_{k,q'}\delta_{q,k'}n_{k}(1-n_{q})\\
\big\langle\hat{d}_{k}\hat{d}_{q}\hat{d}_{k'}^{\dagger}\hat{d}_{q'}^{\dagger}\big\rangle&=\left (-\delta_{k,k'}\delta_{q,q'}+\delta_{k,q'}\delta_{q,k'}\right )(1-n_{k})(1-n_{q})\\
\big\langle\hat{d}_{k}^{\dagger}\hat{d}_{q}^{\dagger}\hat{d}_{k'}\hat{d}_{q'}\big\rangle&=\left (-\delta_{k,k'}\delta_{q,q'}+\delta_{k,q'}\delta_{q,k'}\right )n_{k}n_{q}.\\
\end{split}
\end{equation}
Next, we develop Eq.~\eqref{ter_0} term by term and insert therein the matrix elements as expressed by Eqs.~\eqref{nnt2}. Exchanging the final active labels $k$ and $q$ in the resulting equation, and noting that $n_k^2=n_k$, the compact expression for the TTC $\mathcal{C}_{i,j}\pap{t}$ written as Eq.(21) in the main text follows immediately.

\section{Two-time correlations: analytical results}
Our theoretical work allows us to obtain an exact and closed analytical expression for $\mathcal{C}_{1}^{(x)}\pap{t}$, which is valid for any time and chemical potential. In order to guide the reader, we first evaluate the two-time correlation as
\begin{equation}
\mathcal{C}_{1}^{(x)}\pap{t}= \frac{1}{2}\big\langle \lbrace \hat{\gamma}_1\pap{t},\hat{\gamma}_1 \rbrace\big\rangle
\end{equation}
The temporal evolution for the Majorana operator is given by $\hat{\gamma}\pap{t}=\ee^{\ii \hat{H} t} \hat{\gamma}_{1} \ee^{-\ii \hat{H} t}$, where $\hat{H}$ is the Kitaev-Majorana Hamiltonian Eq.~\eqref{KMH}. We calculate this evolution using the traditional Baker-Campell-Hausdorff formula (BCH) $\ee^{s\hat{A}}\hat{B}\ee^{-s\hat{A}}=\hat{B}+s[\hat{A},\hat{B}]+\frac{s^2}{2!}[\hat{A},[\hat{A},\hat{B}]]+...$, and by a direct substitution we evaluate each commutator as follows:
\begin{equation}
\begin{split}
[\hat{H},\hat{\gamma}_1]&=\ii\mu\,\hat{\gamma}_2\\
[\hat{H},[\hat{H},\hat{\gamma}_1]]&=\mu^2\,\hat{\gamma}_1+2\mu \Delta\, \hat{\gamma}_3\\
[\hat{H},[\hat{H},[\hat{H},\hat{\gamma}_1]]]&=\ii\mu\left ( \mu^2+4\Delta^2 \right )\,\hat{\gamma}_2+2\ii\mu^2 \Delta\, \hat{\gamma}_4\\
[\hat{H},[\hat{H},[\hat{H},[\hat{H},\hat{\gamma}_1]]]]&=\mu^2 \left ( \mu^2+4\Delta^2 \right )\,\hat{\gamma}_1+4\mu\Delta\left (\mu^2+2\Delta^2 \right )\,\hat{\gamma}_3+4\mu^2 \Delta^2\,\hat{\gamma}_5\\
[\hat{H},[\hat{H},[\hat{H},[\hat{H},[\hat{H},\hat{\gamma}_1]]]]]&=\ii\mu \left ( \mu^4+3\mu^2 4\Delta^2+16\Delta^4 \right )\,\hat{\gamma}_2+4\ii\mu^2 \Delta\left ( \mu^2+4\Delta^2 \right )\,\hat{\gamma}_4+4\ii\mu^3 \Delta^2\,\hat{\gamma}_6\\
[\hat{H},[\hat{H},[\hat{H},[\hat{H},[\hat{H},[\hat{H},\hat{\gamma}_1]]]]]]&=\mu^2 \left ( \mu^4+12\mu^2 \Delta^2+16\Delta^4 \right )\,\hat{\gamma}_1+2\mu\Delta\left ( 3\mu^4+5\mu^2 4\Delta^2+16\Delta^4 \right )\,\hat{\gamma}_3\\
&\quad+4\mu^2 \Delta^2\left ( 3\mu^2+8\Delta^2 \right )\,\hat{\gamma}_5+8\mu^3 \Delta^3\,\hat{\gamma}_7.
\end{split}
\end{equation}
After evaluating these first terms of the BCH formula, we can figure out the sequence of the emerging terms. In addition we use the fact that the Majorana operators satisfy the property $\pap{\gamma_j}^{2}=\pap{\gamma_j^{\dagger}}^{2}=\hat{1}$, and that they obey the modified anticommutation relations $\lbrace \gamma_i,\gamma_j\rbrace=2\delta_{i,j}$ with $i,j=1,...,2N$. Therefore, after a careful algebraic process, we find that
\begin{equation}\label{C1_X}
\mathcal{C}_{1}^{(x)}\pap{t}=\sum_{m=0}^{\infty}\frac{(-1)^m}{(2m)!}\left ( 2\Delta t \right )^{2m}\mathcal{N}_m(u^2),
\end{equation}
where $u=\mu/2\Delta$ and $\mathcal{N}_{m}(x)$ are the well-known Narayana polynomials (NP). The NP have the form
\begin{equation}
\mathcal{N}_{m}\pap{x} =\sum_{n=1}^{m}N_{m,n}x^{n}\; \qquad \text{with}\,\qquad N_{m,n}=\frac{1}{m}\binom {m}{n-1}\binom {m}{n}
\end{equation}
where $N_{m,n}$ are denoted as the Narayana numbers.

\subsection{$\mathcal{C}_{1}^{(x)}$-TTC for the critical point $\mu=2\Delta$}
Given the closed form of $C_{1}^{(x)}\pap{t}$ for any chemical potential, we evaluate Eq.~\eqref{C1_X} directly for $u=1$, and note that $\mathcal{N}_m(1)=C_m$ (the $m$th Catalan number), with
\begin{equation}\label{cm1}
C_m=\frac{1}{m+1}\binom {2m}{m}.
\end{equation}
We substitute Eq.~\eqref{cm1} into Eq.~\eqref{C1_X}, and obtain the exact result for $C_{1}^{(x)}\pap{t}$ at the critical point,
\begin{equation}
C_{1}^{(x)}\pap{t}=\frac{\mathcal{J}_1\left ( 4\Delta t \right )}{2\Delta t},
\end{equation}
where $\mathcal{J}_{1}\pap{z}$ is the Bessel Function of the first kind.

\end{document}